\documentclass[letterpaper,english,aps,reprint,superscriptaddress]{revtex4-1}
\usepackage{mathptmx}
\usepackage[T1]{fontenc}
\usepackage[latin9]{inputenc}
\usepackage{color}
\usepackage{amsmath}
\usepackage{amssymb}
\usepackage{graphicx}

\makeatletter

\pdfpageheight\paperheight
\pdfpagewidth\paperwidth

\renewcommand{\fnum@figure}{\text{FIG.~\thefigure}}

\makeatother

\usepackage{babel}
\begin{document}

\title{Probing short-range magnetic order in a geometrically frustrated
magnet by spin Seebeck effect}

\author{Changjiang Liu}

\thanks{These authors contributed equally to this work.}

\affiliation{Materials Science Division, Argonne National Laboratory, Lemont,
Illinois 60439, USA}

\author{Stephen M. Wu}

\thanks{These authors contributed equally to this work.}

\affiliation{Materials Science Division, Argonne National Laboratory, Lemont,
Illinois 60439, USA}

\affiliation{Department of Electrical and Computer Engineering, University of
Rochester, Rochester, New York 14627, USA}

\author{John E. Pearson}

\affiliation{Materials Science Division, Argonne National Laboratory, Lemont,
Illinois 60439, USA}

\author{J. Samuel Jiang}

\affiliation{Materials Science Division, Argonne National Laboratory, Lemont,
Illinois 60439, USA}

\author{N. d\textquoteright Ambrumenil}

\affiliation{Department of Physics, University of Warwick, Coventry CV4 7AL, UK}

\author{Anand Bhattacharya}
\email{anand@anl.gov}

\affiliation{Materials Science Division, Argonne National Laboratory, Lemont,
Illinois 60439, USA}
\begin{abstract}
Competing magnetic interactions in geometrically frustrated magnets
give rise to new forms of correlated matter, such as spin liquids
and spin ices. Characterizing the magnetic structure of these states
has been difficult due to the absence of long-range order. Here, we
demonstrate that the spin Seebeck effect (SSE) is a sensitive probe
of magnetic short-range order (SRO) in geometrically frustrated magnets.
In low temperature (2 - 5 K) SSE measurements on a model frustrated
magnet $\mathrm{Gd_{3}Ga_{5}O_{12}}$, we observe modulations in the
spin current on top of a smooth background. By comparing to existing
neutron diffraction data, we find that these modulations arise from
field-induced magnetic ordering that is short-range in nature. The
observed SRO is anisotropic with the direction of applied field, which
is verified by theoretical calculation. 
\end{abstract}
\maketitle
Pure spin currents carried by magnetic excitations are of fundamental
interest and may be used to transmit and store information \cite{Bauer2012}.
One method of generating a pure spin current is through the spin Seebeck
effect (SSE), where a thermal gradient drives a current of magnons.
Spin currents have been generated in this way using both ferromagnetic
(FM) \cite{Uchida2010} and antiferromagnetic (AFM) \cite{Wu2016,Seki2015}
magnons. It has been shown that for correlated paramagnetic insulators,
a spin current may be generated via the SSE or paramagnetic spin pumping
\cite{Shiomi2014,Wu2015}. It is presumed that this is due to short
lived magnons (paramagnons) arising as a result of correlations between
spins \cite{Fleury1969,Mila1991,Doubble2010,Shiomi2014,Okamoto2016,Qin2017}. 

Current understanding of the SSE in magnetic insulators is based on
the diffusion of thermally activated magnons \cite{Xiao2010,Adachi2013,Rezende2016,Rezende2016a}.
Such a mechanism is supported by recent experiments studying the length
scale, temperature and magnetic field dependencies of SSE in ferrimagnetic
insulator Yttrium Iron Garnet (YIG) \cite{Kehlberger2015,Uchida2014,Ritzmann2015}.
The diffusive magnons have finite lifetime and diffusion length. The
fact that SSE can be measured in nanometer thickness YIG films and
in the picosecond time scale \cite{Kimling2017} suggests that the
SSE is sensitive to magnons in very small volume or with very short
lifetimes. These aspects of SSE suggest that it may be used as a sensitive
probe of magnetic order in unconventional magnetic materials, such
as geometrically frustrated systems. 

\begin{figure}[h]
\begin{centering}
\includegraphics{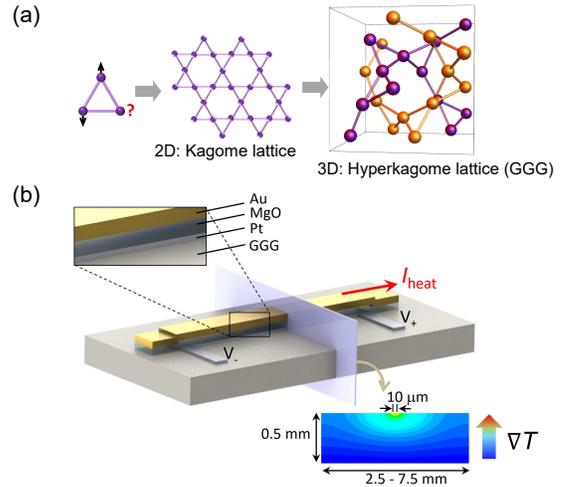}
\par\end{centering}
\caption{GGG hyperkagome lattice and schematics of the SSE device. (a) Illustration
of geometrical frustration of antiferromagnetically coupled spins
on a triangular lattice. Kagome lattice is a realization of such frustration
in two dimensional space. When extended to three dimensions, corner-sharing
triangles form the hyperkagome lattice. For GGG, the two interpenetrating
corner-sharing triangular sublattices are shown in purple and orange,
respectively. (b) Device design of an on-chip heated SSE device. Upper
left panel shows the vertical structure of the fabricated device.
A cross-sectional view of the simulated temperature profile in GGG
is shown in the bottom panel. }
\end{figure}

In this work, we use the SSE to probe the magnetic short-range order
(SRO) in a model frustrated magnet gadolinium gallium garnet ($\mathrm{Gd}{}_{3}\mathrm{Ga}{}_{5}\mathrm{O}{}_{12}$,
GGG). It was shown earlier that the generation of a spin current
by SSE does not need magnetic long-range order (LRO) \cite{Wu2015,Hirobe2017}.
While this suggests that spin correlations or SRO may play a role,
a definitive demonstration of a connection between SRO and the SSE
has been missing. Here, we demonstrate for the first time that a specific
antiferromagnetic order that is short-range in nature can be detected
by the SSE in the absence of LRO.

As is shown in Fig. 1(a), the Gd sites in GGG form a hyperkagome lattice,
a three-dimensional kagome lattice consisting of two interpenetrating
corner-sharing triangular sublattices \cite{Petrenko1998,Deen2015}.
The exchange interaction between nearest-neighbor $\mathrm{Gd}^{3+}$
ions are antiferromagnetic. Owing to geometrical frustration on the
hyperkagome lattice, GGG hosts an rich phase diagram at low temperatures
($T<1$ K) \cite{Deen2015}. There is no magnetic LRO in GGG down
to 25 mK, even though the Curie-Weiss temperature is $\theta_{\mathrm{CW}}\sim-2.3$
K \cite{Kinney1979,Petrenko1998,Dunsiger2000,Wu2015}. It has been
shown that many interesting phases arise within GGG, including spin
liquid states \cite{Petrenko1998}, protected spin clusters \cite{Ghosh2008}
and a hidden multipolar order \cite{Paddison2015}. In our experiment,
we use the SSE to probe magnetic-field-induced SRO in GGG in the temperature
regime (2 - 5 K) where effects due to geometric frustration starts
to emerge as we approach the magnitude of $\theta_{\mathrm{CW}}$.

\begin{figure}[h]
\begin{centering}
\includegraphics{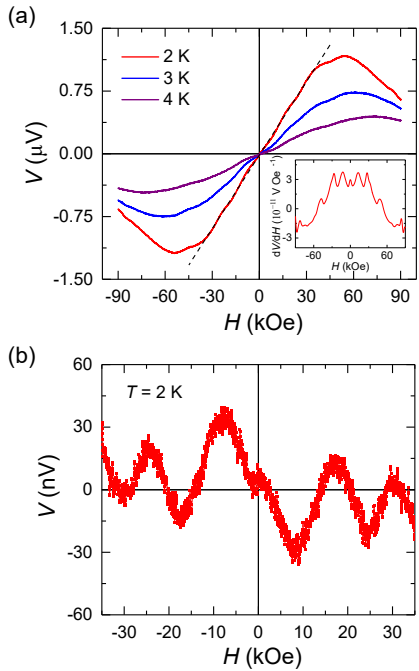}
\par\end{centering}
\caption{SSE measurement at low temperatures and modulation in the SSE response.
(a) SSE measurement results at temperatures below 5 K. Inset is the
derivative of SSE voltage with respect to magnetic field for $T$
= 2 K data. (b) Field-dependent modulation in the SSE response is
seen after subtracting a linear background (dashed line in Fig. 2(a))
from the SSE signal in the field range - 35 kOe $<H<$ 35 kOe.}
\end{figure}

To accomplish this, SSE devices were patterned onto GGG single crystals
with polished surface along (111) or (001) (see also Supplemental
Material). Platinum (Pt) was used as spin detector material. Local
heating was achieved by passing an electric current through a gold
heater wire \cite{Wu2015a}, electrically isolated from the spin detector
layer by a thin MgO layer. The resulting temperature gradient is perpendicular
to the sample plane, which drives spin excitations from the GGG into
the Pt detector, where a voltage develops as a result of the inverse
spin Hall effect (ISHE). Figure 1(b) depicts the structure of the
fabricated SSE device. Using these devices, our measurement results
agree with the original SSE experiments on GGG \cite{Wu2015} which
was carried out at a higher range of temperature ($T$ > 5 K). Shown
in Fig. 2(a) are the SSE signals measured as a function of magnetic
field at different temperatures. The initial rise of SSE response
with magnetic field is due to the increase of magnetization of GGG
until almost saturation. However, we observe an appreciable downturn
in the SSE response at lower temperatures in the high field range,
as can be seen in the $T=$ 2 K data. This downturn is presumably
caused by the opening of a Zeeman gap in the magnon spectrum, similar
to the observation in ferrimagnetic insulator YIG \cite{Kikkawa2015,Ritzmann2015,Guo2016}.
To confirm this, we performed the SSE measurement down to 200 mK,
where the SSE signal is suppressed to zero for $H>90$ kOe (see Supplementary
Fig. 1).

As described above, GGG is not simply a paramagnet. It possesses strong
geometric frustration due to antiferromagnetic and dipolar interactions
on the hyperkagome lattice. If one considers only the nearest-neighbor
antiferromagnetic exchange interaction $J$, it has been shown that
for a hyperkagome lattice a magnetic field with energy scale equal
to 6 $J$ or about 17 kOe is required to align those spins in GGG
even at $T=0$ K \cite{Zhitomirsky2000}. Our SSE experiment exhibits
this behaviors in that the maximum signal (around saturation) occurs
at fields much higher than the value inferred by a Brillouin function
for a non-interacting paramagnet at the same low temperature (\textcolor{black}{Supplementary
Fig. }2). We note that (and will be discussed later) because of the
large dipolar interactions, the magnetic moments in GGG are not fully
aligned even at fields beyond 17 kOe. 

Upon closer examination, the SSE signals shown in Fig. 2(a) are found
not to be smooth functions of magnetic field, but contain considerable
field-dependent modulations on top of the S-shaped curve. Inset of
Fig. 2(a) shows the derivative of the SSE signal with respect to magnetic
field, in which we can clearly see the modulation in the SSE response
as a function of field. In Fig. 2(b), the modulation in SSE voltage
is plotted directly after subtracting a linear background (indicated
by the dashed line in Fig. 2(a)) from the SSE signal in the field
range - 35 kOe $<H<$ 35 kOe. When the same experiment is performed
on a single crystal YIG sample in the same temperature and magnetic
field range, the modulation is absent (Supplementary Fig. 3). We also
rule out the possibility of magnetic-field-induced thermal conductivity
or heat capacity changes as a possible source for this effect by independently
measuring these quantities as a function of applied magnetic field
(Supplementary Fig. 4).

\begin{figure}[h]
\begin{centering}
\includegraphics{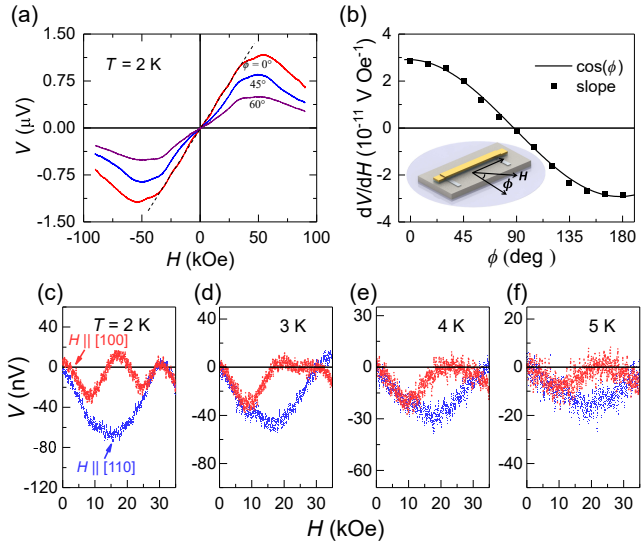}
\par\end{centering}
\caption{Anisotropic behavior of the modulation in SSE response. (a) SSE signals
measured with magnetic field applied along different directions in
the (001) plane, with $\phi=0\protect\textdegree$ being along {[}100{]}
crystal axis. (b) Magnitude of slope of the linear part of the SSE
signal satisfies cosine angular dependence. Inset shows the corresponding
measurement geometry. $H$ is the applied in-plane magnetic field.
(c) to (f) Modulations in SSE voltage from T = 2 to 5 K. Data in red
and blue corresponds to field applied along {[}100{]} and {[}110{]}
directions, respectively. In these measurement, the SSE devices were
fabricated on the (001) surface of GGG signal crystals.}
\end{figure}

In neutron scattering studies on GGG, it has been found that different
magnetic orders rise and fall with increasing magnetic field at low
temperature (< 400 mK) \cite{Schiffer1994,Petrenko1999,Petrenko2002,Petrenko2009,Deen2015}.
At higher temperatures ($T\sim3$ K, as is in our measurement range),
short-range correlations are known to persist \cite{Petrenko1998,Petrenko2000},
though little is known about their properties under a magnetic field.
The dynamics of thermally excited magnons in this temperature range
depends on these short-range correlations, which in turn can influence
the SSE signal. It is also known from neutron scattering and bulk
magnetometry measurements that the field-induced magnetic orderings
in GGG have distinct anisotropies. For instance, the critical field
at which the AFM phase emerges \cite{Petrenko2002,Petrenko2009,Deen2015}
depends on the direction of the field relative to the GGG crystal
axis, being different for field aligned along {[}100{]} versus along
{[}110{]} crystal axis \cite{Deen2015,Rousseau2017}. Such anisotropy
is presumably caused by dipolar interactions among Gd ions \cite{2015}.
The $\mathrm{Gd}^{3+}$ ion in GGG carries a relatively large magnetic
moment of 7 $\mu_{B}$ leading to the dipolar interactions with an
energy scale comparable to the nearest-neighbor exchange interaction. 

To find out whether the modulations in the SSE signal are associated
with a field-induced magnetic ordering, we performed the SSE measurements
with magnetic field applied along several different GGG crystal axes.
The measured SSE signals from the same device are shown in Fig. 3(a),
with 0 and 45 degree corresponding to {[}100{]} and {[}110{]} crystal
axes, respectively. Figure 3(b) shows the overall magnitude of the
SSE signal, represented by the slope of the linear background at low
fields, as a function of the in-plane angle of the magnetic field.\textcolor{black}{{}
The data is fit well by a cosine function which is expected from the
geometry of the ISHE, and implies that the linear background is isotropic
and independent of the direction of magnetic field. In Fig. 3(c) -
(f), the modulations in the SSE signal are presented for temperatures
from 2 to 5 K.} In these plots, red and blue data correspond to field
applied along {[}100{]} and {[}110{]} GGG crystal axes, respectively.
For clarity, only data for the positive field range is plotted. The
magnitude of these modulations becomes smaller as temperature increases.
At each temperature, the magnetic-field dependences of the modulation
are clearly different between the two field orientations. For instance,
in Fig. 3(c), when the field is applied along {[}100{]} direction,
the modulation initially decreases with magnetic field and reaches
a minimum at $H\sim8$ kOe, while the minimum for field parallel to
{[}110{]} direction occurs at $H\sim17$ kOe. This same trend is observed
at higher temperatures up to 5 K. At $T=2$ K, however, there is a
second minimum at higher field $H\sim24.5$ kOe for field along {[}100{]}
direction, which is suppressed at $T>3$ K suggesting that this may
have a different origin than the modulations at lower fields.

\begin{figure}[h]
\begin{centering}
\includegraphics{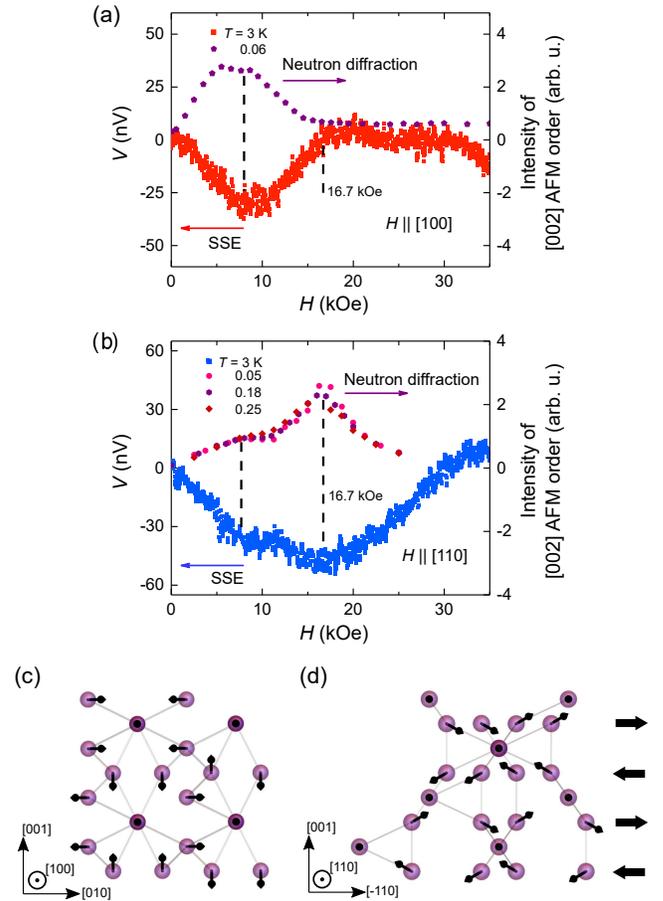}
\par\end{centering}
\caption{Comparison between the SSE response and the intensity of {[}002{]}
AFM order, and theoretical calculation of spin configurations. (a)
and (b) Magnetic field dependencies of the modulation in SSE signal
and the intensity of {[}002{]} AFM order measured by neutron diffraction
with corresponding field applied along {[}100{]} and {[}110{]} crystal
axes, respectively. Vertical dashed lines in both (a) and (b) indicate
the same peak positions in the SSE data and neutron diffraction results.
(c) and (d) Calculated spin configurations in one GGG primitive cell
at an applied field $H=17$ kOe aligned along {[}100{]} and {[}110{]}
crystal axes, respectively. The magnetic field points out of the  page,
and the spin orientation of $\mathrm{Gd^{3+}}$ ions is represented
by an small arrow at each Gd site. Notice that most spins are canted
away from the applied field. In (c) the net spin components in horizontal
direction are zero, while the total horizontal components in (d),
indicated by large arrows on the right, form an alternating AFM pattern. }
\end{figure}

Magnetic ordering in GGG has been extensively studied over the last
two decades, and a summary of these results is presented in the Supplementary
Material. Upon application of a magnetic field at low temperatures
there are FM, AFM and incommensurate AFM LRO that emerge in GGG. Most
of the AFM orderings, including all incommensurate ones, have a strong
temperature dependence, and they are suppressed at $T>400$ mK \cite{Deen2015}.
However, of particular interest is the {[}002{]} AFM order, whose
intensity is observable up to 900 mK \cite{Petrenko2002,Petrenko2009},
above which there is no published experimental data, though LRO is
completely suppressed by 1.3 K (see Supplementary Fig. 5). Shown in
Fig. 4(a) ( reproduced from Ref. \cite{Petrenko2002}) are the integrated
intensity of the {[}002{]} AFM peak as a function of magnetic field
with the field applied along {[}100{]} direction. We see that the
maximum position of the neutron scattering data matches the field,
indicated by the vertical dashed line, where the modulation in SSE
voltage shows a minimum. Crucially, for field applied along {[}110{]}
direction, the {[}002{]} AFM order reaches its maximum at $H\sim16.7$
kOe as shown in Fig. 4(b) (reproduced from Ref. \cite{Petrenko2009})
which is also in good agreement with corresponding SSE measurement
results. Additionally, a small inflection in the SSE data for $H\parallel[110]$
can be seen at $H\sim7.7$ kOe, that is also present in the neutron
diffraction result. Over the entire temperature range available, the
peak positions in neutron scattering data and the minima of our SSE
modulations show little temperature dependence. Any remaining discrepancies
may be due to non-uniform crystal shape in the neutron diffraction
experiments where the demagnetization field is not considered. These
comparisons provide strong evidence that the modulation observed in
the SSE measurement is associated with the {[}002{]} AFM order. The
temperatures used in the SSE measurement are much higher than that
for the disappearance of LRO peaks as measured in neutron scattering
experiments \cite{Deen2015} (see also Supplementary Fig. 5) implying
that this magnetic order is short range in character. 

In order to understand the anisotropic behaviors observed in the SSE
experiment, we have computed the spin order in GGG in the presence
of a magnetic field using a Hamiltonian including exchange ($J$)
and dipolar ($D$) interactions (see also Supplementary Note):
\begin{equation}
\begin{aligned}H & =\underset{j\alpha,l\beta}{\sum}J_{j\alpha,l\beta}\mathbf{S}_{j\alpha}\cdot\mathbf{S}_{l\beta}-g\mu_{B}H\underset{j\alpha}{\sum}S_{j\alpha}^{z}\\
 & +D\underset{j\alpha,l\beta}{\sum}(\frac{\mathbf{S}_{j\alpha}\cdot\mathbf{S}_{l\beta}-3(\mathbf{S}_{j\alpha}\cdot\hat{r}_{j\alpha l\beta})(\mathbf{S}_{l\beta}\cdot\hat{r}_{j\alpha l\beta})}{r_{j\alpha l\beta}^{3}}),
\end{aligned}
\end{equation}
where vectors $\mathbf{S}$ represent the spins at each Gd site with
indices $j$, $l$ identifying the unit cell, and $\alpha$, $\beta$
indicating the twelve Gd ions in the primitive cell, respectively.
The second term in the Hamiltonian comes from the applied magnetic
field $H$, with $g$ and $\mu_{B}$ being the g-factor of Gd ions
and Bohr magneton, respectively. The vectors $\hat{r}$ denote unit
vectors along the direction from site $j\alpha$ to site $l\beta.$
An example of the calculated spin configurations are shown in Fig.
4(c) and (d), with $H=$ 17 kOe applied along {[}100{]} and {[}110{]}
GGG crystal axes, respectively. In each graph, the perspective is
chosen such that the magnetic field points out of the page. In both
cases, spins at most of the Gd sites are canted away (showing transverse
components) from the applied field as a result of the exchange and
dipolar interactions. At $H\sim17$ kOe, the modulation in SSE signal
is almost zero for field along {[}100{]} ( Fig. 4(a)), while its magnitude
becomes largest for field along {[}110{]} (Fig. 4(b)),. Correspondingly,
our calculation of the spin configuration shows that the magnetic
ordering is different between the two cases. For field along {[}110{]},
we find that the canting of spins leads to a net magnetization within
each layer (Fig. 4(d)), with AFM ordering between layers at the {[}002{]}
wavevector, as is also observed in neutron scattering measurement.
In contrast, when the same field is applied along {[}100{]} (Fig.
4(c)), no AFM order develops.

In general, the SSE is associated with two effects, the magnetization
carried by thermally excited magnons and their diffusivity. According
to our calculation, the total magnetization of GGG at $H=17$ kOe,
taking into account the canting of spins, is very similar for the
two field directions. This suggests that the large isotropic \textquoteleft background\textquoteright{}
SSE signal may originate from excitations derived from the total magnetization.
In contrast, the magnetic order due to the canting of spins is different for the two field directions. This suggests that the modulations in
the SSE signal, which decreases as AFM order increases, could be due to a decrease in the number of magnons excited since AFM ordering may lead to a gap in the spin-wave excitation spectrum \cite{Quilliam2007}. The decrease in SSE could also be due
to changes in the spin-wave dispersion with the onset of AFM order that lower the diffusivity of magnons.

In conclusion, we have demonstrated that the spin Seebeck effect,
in addition to serving as a generator of spin current for spintronics
applications, can also be used as a more general technique to probe
magnetic order in a larger class of condensed matter systems, such
as geometrically frustrated magnets studied in this work. For gadolinium
gallium garnet, our SSE measurements have revealed a field-induced,
anisotropic short-range order that persists to high temperatures (up
to 5 K), which was not known previously. This new approach, where
we use SSE to probe magnetic structures in the absence of long-range
order, opens the door to exploring frustrated quantum magnetic systems,
particularly for samples with limited volume such as exfoliated materials
and thin films. This would allow us to probe collective excitations
that are \textit{only} short-ranged in nature, and thus entirely hidden
to the community, and serve as a guide for large-scale neutron or
x-ray scattering experiments. 
\begin{acknowledgments}
We acknowledge very valuable discussions with Oleg Petrenko and thank
him for sharing neutron data on GGG. All work at Argonne was supported
by the U.S. Department of Energy, Office of Science, Basic Energy
Sciences, Materials Sciences and Engineering Division. The use of
facilities at the Center for Nanoscale Materials, an Office of Science
user facility, was supported by the U.S. Department of Energy, Basic
Energy Sciences under contract No. DEAC02-06CH11357.
\end{acknowledgments}

\end{document}